\documentclass[11pt,twoside]{article}

\usepackage{asp2014}

\aspSuppressVolSlug
\resetcounters

\begin{document}

\title{Multi-segment and Echelle stellar spectra processing issues and how to solve them}

\author{Sviatoslav Borisov,$^{1,2}$ Igor Chilingarian,$^{3,2}$ Evgenii Rubtsov,$^{2,4}$ Kirill Grishin,$^{5,2}$ Ivan Katkov,$^{2,6,7}$ Vladimir Goradzhanov,$^{2,4}$ Anton Afanasiev,$^{5,2}$ Anna Saburova,$^2$ Anastasia Kasparova,$^2$ and Ivan Zolotukhin$^2$}
\affil{$^1$Department of Astronomy, University of Geneva, Versoix, Switzerland; \email{sviatoslav.borisov@unige.ch}}
\affil{$^2$Sternberg Astronomical Institute, M.V. Lomonosov Moscow State University, Moscow, Russia}
\affil{$^3$Center for Astrophysics -- Harvard and Smithsonian, Cambridge, MA, USA}
\affil{$^4$Faculty of Physics, Moscow State University, Moscow, Russia}
\affil{$^5$Universit\'e de Paris, CNRS, Astroparticule et Cosmologie, Paris, France}
\affil{$^6$New York University Abu Dhabi, UAE}
\affil{$^7$Center for Astro, Particle, and Planetary Physics, NYU AD, UAE}

\paperauthor{Borisov~Sviatoslav}{sviatoslav.borisov@unige.ch}{0000-0002-2516-9000}{University of Geneva}{Department of Astronomy}{Versoix}{Canton of Geneva}{1290}{Switzerland}
\paperauthor{Igor Chilingarian}{igor.chilingarian@cfa.harvard.edu}{0000-0002-7924-3253}{Center for Astrophysics - Harvard and Smithsonian}{}{Cambridge}{}{02138}{USA}
\paperauthor{Evgenii Rubtsov}{rubtsov602@gmail.com}{0000-0001-8427-0240}{Sternberg Astronomical Institute, Lomonosov Moscow State University}{}{Moscow}{}{119234}{Russia}
\paperauthor{Kirill Grishin}{kirillg6@gmail.com}{0000-0003-3255-7340}{Sternberg Astronomical Institute, Lomonosov Moscow State University}{}{Moscow}{}{119234}{Russia}
\paperauthor{Ivan Katkov}{katkov.ivan@gmail.com}{0000-0002-6425-6879}{NYU Abu Dhabi}{Center for Astro, Particle, and Planetary Physics}{Abu Dhabi}{}{129188}{UAE}
\paperauthor{Vladimir Goradzhanov}{goradzhanov.vs17@physics.msu.ru}{0000-0002-2550-2520}{Sternberg Astronomical Institute, Lomonosov Moscow State University}{}{Moscow}{}{119234}{Russia}
\paperauthor{Anton Afanasiev}{afanasa25@gmail.com}{0000-0002-8220-0756}{Sternberg Astronomical Institute, Lomonosov Moscow State University}{}{Moscow}{}{119234}{Russia}
\paperauthor{Anna Saburova}{saburovaann@gmail.com}{0000-0002-4342-9312}{Sternberg Astronomical Institute, Lomonosov Moscow State University}{}{Moscow}{}{119234}{Russia}
\paperauthor{Anastasia Kasparova}{anastasya.kasparova@gmail.com}{0000-0002-1091-5146}{Sternberg Astronomical Institute, Lomonosov Moscow State University}{}{Moscow}{}{119234}{Russia}
\paperauthor{Ivan Zolotukhin}{ivan.zolotukhin@gmail.com}{0000-0002-5544-9476}{Sternberg Astronomical Institute, Lomonosov Moscow State University}{}{Moscow}{}{119234}{Russia}


\begin{abstract}
High quality stellar spectra are in great demand now -- they are the most important ingredient in the stellar population synthesis to study galaxies and star clusters. Here we describe the procedures to increase the quality of flux calibration of stellar spectra. We use examples of NIR intermediate-resolution Echelle spectra collected with the Folded InfraRed Echellete (R$\sim$6500, Magellan Baade) and high-resolution UV--optical spectra observed with UVES (R$\sim$80000, ESO VLT). By using these procedures, we achieved the quality of the global spectrophotometric calibration as good as 1--2\%, which fulfills the requirements for the quality of stellar spectra intended to be used in the stellar population synthesis.
\end{abstract}



\section{Introduction}
Modern astrophysics faces new challenges that require high-quality spectral data. Stellar spectra are a crucial component that helps us to understand properties of individual stars as well as stellar populations in star clusters and galaxies. A significant fraction of state-of-the-art spectrographs are cross-dispersed Echelle and/or have multiple setups (segments), which are combined together. This approach covers a longer wavelength range at high spectral resolution but makes data reduction and post-processing challenging. Here we use examples of spectra from the UVES-POP stellar spectral library \citep{Bagnulo2003} observed with UVES \citep{Dekker2000} and NIR Echelle spectra \citep{Chilingarian2015} observed with Folded InfraRed Echellete (FIRE, \citealp{Simcoe2008})  and describe how to bring multi-order/multi-segment spectra to perfection by identifying various issues that one could face when dealing with them and approaches for solving these issues. 

\section{On the issues in processing of multi-segment and Echelle stellar spectra}
Echelle orders might end up systematically wrong fluxes after extraction leading to ripples in the regions where the orders overlap. This is caused by the offset of a science frame with respect to flat field exposures amplified by the optics and  under- or over-subtraction of scattered light. Probable reasons for the offsets of science frames are mechanical flexures and significant temperature/pressure changes in the dome. 

Another procedure which might be problematic is a telluric correction. The standard way to perform it is to use spectra of so-called telluric standards \citep{Vacca2003}, that is ``featureless'' A0V stars, however, they are not ideal because of different rotational velocities and probable presence of emission lines in their spectra. Moreover, there might be a mismatch of real stellar atmospheric parameters of a telluric star and its model parameters. Also, because telluric absorption lines are very narrow, even small wavelength calibration errors lead to very significant residuals. 

The quality of flux calibration can also be affected by imperfection of sensitivity curves determined by the observations of spectrophotometric standards. It might be caused by various issues (e.g. changes in observing conditions, transparency, etc.) or imperfect reference spectra of the ``standard'' stars used for the throughput determination or their stellar atmospheric models. 

\section{Data reduction approaches}
\begin{figure}[h]
\centering
\includegraphics[width=0.7\linewidth]{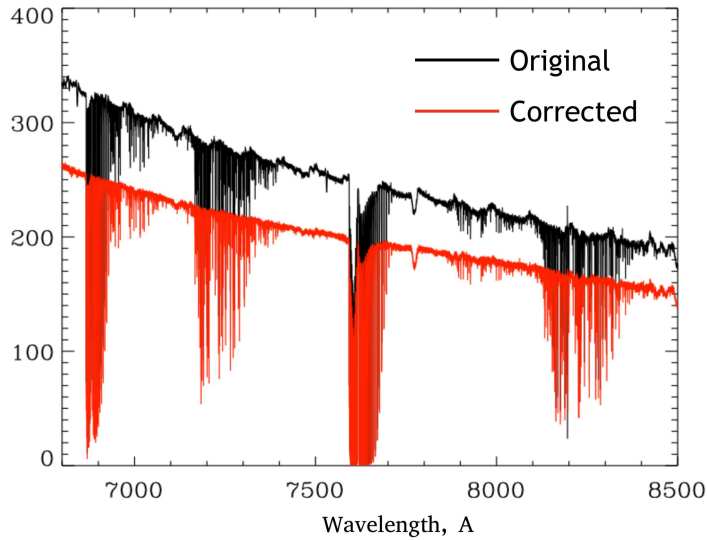}
\caption{Original and ripples-corrected spectra of HD320764 from the UVES-POP library.}
\label{O2-003_ripples}
\end{figure}

For UVES spectra, we developed an algorithm for the accurate stitching of Echelle orders which eliminates ripples in the regions where they overlap; this algorithm has been successfully integrated into the processing sequence consisting of procedures provided by ESO \citep{Borisov2020}. For the FIRE and MagE \citep{Marshall2008} spectrographs operated at the 6.5-m Magellan Baade telescope and ESI operated at the 10-m Keck, we developed a complete data reduction pipeline from scratch \citep{Chilingarian2020}, which implements a similar algorithm prior to stitching of Echelle orders. The key issue is to understand how the general principles of the optical design (telecentricity violations, flexures) can affect the final data product when using standard internal flat field and arc calibration frames. Figure~\ref{O2-003_ripples} demonstrates an example of the correction of ripples.

\begin{figure}[h]
\centering
\includegraphics[width=0.8\linewidth]{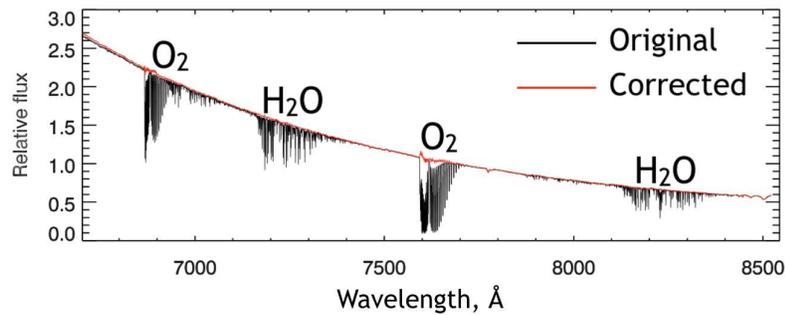}
\caption{Spectrum of HD162305 before (black) and after telluric correction (red).}
\label{O2-003_tell_corr}
\end{figure}

\begin{figure}[ht!]
\centering
\includegraphics[width=0.8\linewidth]{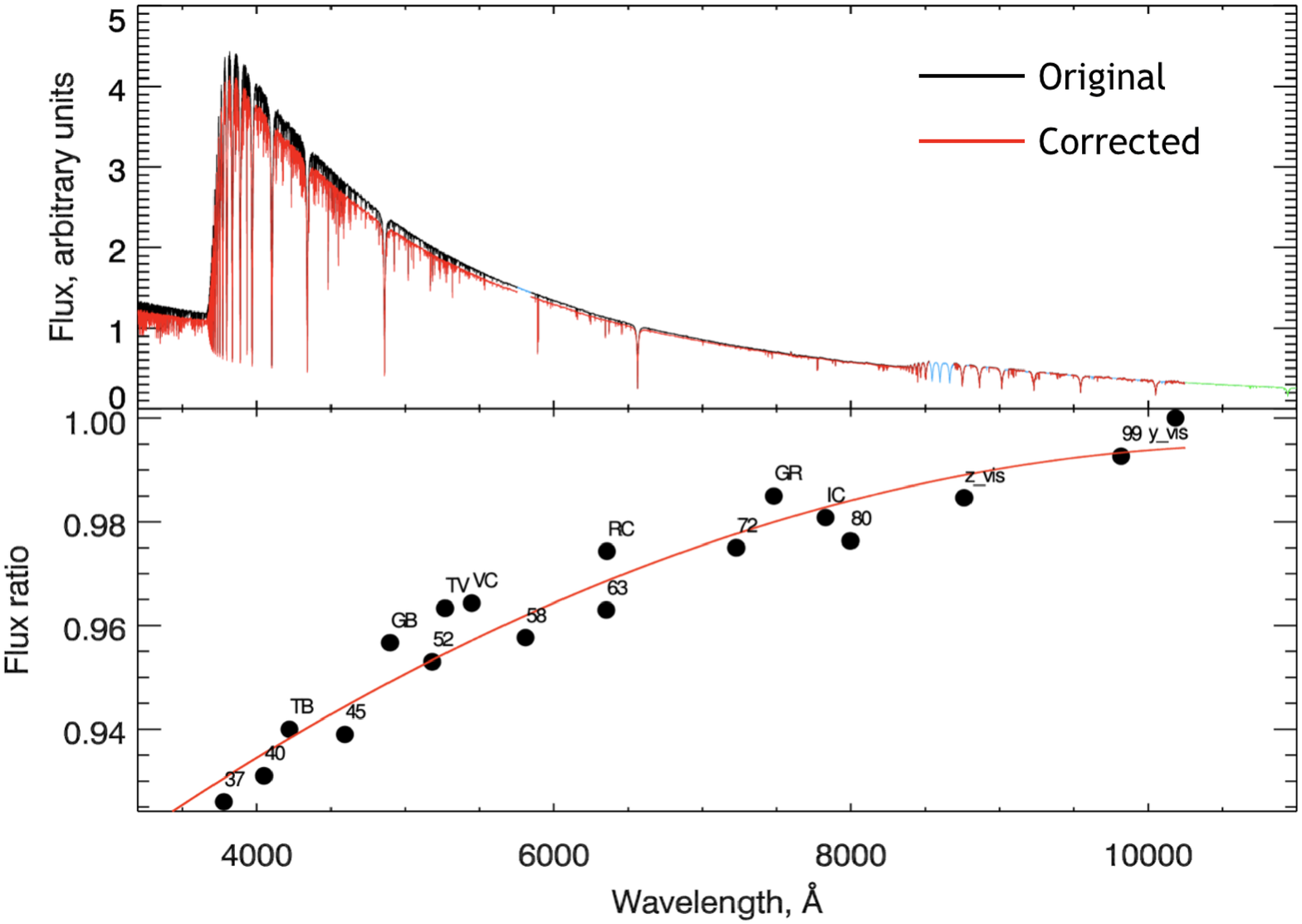}
\caption{Top panel: spectrum of HD75063 from UVES-POP before the spectrophotometric correction (black line) and after (red line). Bottom panel: the points are the ratios of fluxes calculated from observed magnitudes to synthetic fluxes calculated from the spectrum, the  red  line  is  a final correcting  polynomial.}
\label{O2-003_phot_corr}
\end{figure}

We use our own telluric correction procedure that fits an observed spectrum against a linear combination of synthetic stellar spectra multiplied by a pre-computed grid of atmospheric transmission models from {\sc skycalc} \citep{Noll2012} in the airmass-PWV (precipitable water vapor) parameter space taking into account rotational broadening of stellar absorptions (see \citet{Afanasiev2018} and Borisov et al. in prep.). Fig.~\ref{O2-003_tell_corr} shows an example of the telluric correction for HD162305 from the UVES-POP stellar library.

Finally, we use the Virtual Observatory to collect broad- and middle-band photometry, which we then use to perform the global spectral sensitivity correction of final merged spectra. See figure~\ref{O2-003_phot_corr} for an example of spectrophotometric correction.

The quality of the flux correction is illustrated by the fact that a pixel-space full spectrum fitting procedure applied to stellar spectra collected with UVES and FIRE, and galaxy/star cluster spectra collected with MagE does not require a high-order multiplicative continuum correction to match synthetic spectra of stars or stellar populations (see \citealp{X7-003_adassxxxi}).

The spectra from the UVES-POP and LCO-SL stellar libraries are presented on the website of the project (\url{https://sl-dev.voxastro.org/}). It provides access to re-calibrated spectra of stars along with their atmospheric parameters which were computed by using the method developed by our team and based on the minimization technique briefly described in Borisov et al. (in prep.) and \citet{X7-003_adassxxxi}.

\acknowledgements This project is supported by the RScF grant 17-72-20119 and the Interdisciplinary Scientific and Educational School of Moscow University ``Fundamental and Applied Space Research''.

\bibliography{O2-003}  


\end{document}